\documentclass[twocolumn]{aastex7}
\usepackage{natbib,graphics,graphicx,multirow,amsmath,cancel,color,xcolor,hyperref}
\usepackage{soul}
\usepackage[normalem]{ulem}
\usepackage{float}
\usepackage{comment}
\usepackage{tabularx}
\usepackage{multirow}
\usepackage{array}
\usepackage{enumitem}
\renewcommand{\arraystretch}{1.2}
\usepackage{color, colortbl}
\usepackage{bm}
\usepackage[dvipsnames]{xcolor}

\begin{document}
\title{On the Relationship Between Nanoflare Energy and Delay in the Closed Solar Corona}

\correspondingauthor{Shanwlee Sow Mondal}
\email{sowmondal@cua.edu} 
\author[0000-0003-4225-8520]{Shanwlee Sow Mondal}
\affil{Department of Physics, The Catholic University of America, Washington, DC, USA}
\affil{Heliophysics Science Division, NASA Goddard Space Flight Center 8800 Greenbelt Rd., Greenbelt, MD 20771, USA}
\email{sowmondal@cua.edu, shanwlee.sowmondal@nasa.gov, shanwlee.sowmondal@gmail.com}

\author[0000-0003-2255-0305]{James A. Klimchuk}
\affil{Heliophysics Science Division, NASA Goddard Space Flight Center 8800 Greenbelt Rd., Greenbelt, MD 20771, USA}
\email{james.a.klimchuk@nasa.gov}

\author[0000-0003-4023-9887]{C. D. Johnston}
\affil{Physics and Astronomy Department, George Mason University, Fairfax, VA, USA}
\affil{Heliophysics Science Division, NASA Goddard Space Flight Center 8800 Greenbelt Rd., Greenbelt, MD 20771, USA}
\email{cjohn44@gmu.edu}

\author[0000-0002-1198-5138]{L. K. S. Daldorff}
\affil{Department of Physics, The Catholic University of America, Washington, DC, USA}
\affil{Heliophysics Science Division, NASA Goddard Space Flight Center 8800 Greenbelt Rd., Greenbelt, MD 20771, USA}
\email{lars.daldorff@nasa.gov}

\begin{abstract}
Determining the relationship between nanoflare energies and their delays is the key for understanding the physical mechanism of the events and the  plasma response. Nanoflares analyzed in this study were generated self-consistently via prescribed photospheric motions in a 3D multi-strand simulation of a subset of active region magnetic flux. Energies and durations were quantified using three distinct methods. In this study, we investigated the correlation between nanoflare energies (E) and delays ($\tau_D$) using two non-parametric, rank-based statistical tests. Across all methods, results consistently show little to no correlation. This is further supported by the distribution of the exponent $\alpha$ in the assumed relation $E \propto \tau_D^\alpha$, which peaks near zero, and by broad delay distributions within fixed energy bins. These findings are irrespective of whether delays are correlated with the energy of the preceding or subsequent event. They also hold for a subset of high-energy nanoflares. The absence of correlation suggests that nanoflare onset is not solely determined by a critical value of magnetic stress and may involve triggering by other events, perhaps related to a locally complex topology.
 
\end{abstract}

\section{Introduction}
In the absence of continuous energy input, the solar corona would cool rapidly due to radiative and conductive losses \citep{Withbroe_1977ARA&A..15..363W}. A widely accepted theory for maintaining the coronal temperatures in active regions is the nanoflare heating model, in which photospheric motions twist and tangle the magnetic field lines, building up stress in coronal magnetic field. This stress is intermittently released through magnetic reconnection, producing numerous small, impulsive heating events known as nanoflares \citep{Parker_1983, Parker_1988, Klimchuk_2006, Klimchuk_2015}. 

A key diagnostic for understanding this heating mechanism is the correlation between nanoflare energy and the delay between successive nanoflares within a magnetic strand. For example, in the low-frequency nanoflare regime, the time delay between successive nanoflares on a strand exceeds its characteristic cooling time, allowing the plasma to fully cool and drain before being reheated \citep{Bradshaw_2012ApJ...758...53B}. Conversely, in the high-frequency regime, the time between nanoflares is shorter than the cooling time, preventing the strand from fully cooling before the next heating event occurs \citep{Klimchuk_2015}. In the limit of very high-frequency nanoflares, the heating approaches a steady-state condition. Hence, the nanoflare energy-delay relationship provides insight into the plasma response to these heating events as well as help in understanding the underlying mechanism governing magnetic energy storage and release in the corona.

\begin{figure*}
 \centering
 \includegraphics[width=0.99\textwidth,angle=0]{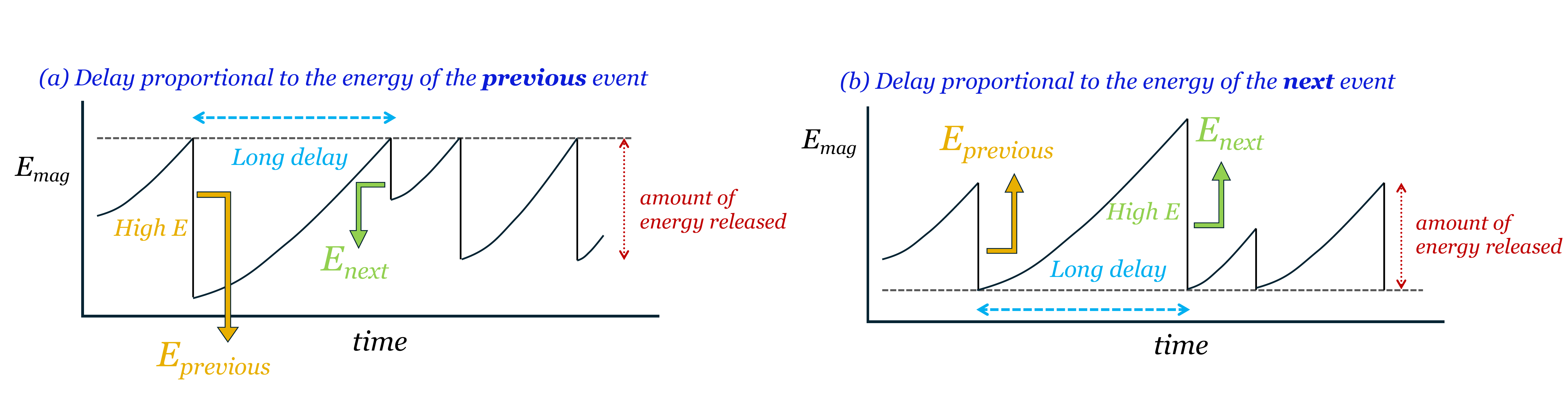}
 \caption{A schematic illustrating two theoretical models of the correlation between nanoflare delay and energy: (a) delay proportional to the energy of the previous event, and (b) delay proportional to the energy of the next event. In panel (a), the dashed black line represents a critical stress threshold (a `ceiling'), beyond which magnetic reconnection is triggered, releasing a portion of the stored free energy. In panel (b), the dashed black line denotes a minimum energy state (a `floor'), suggesting that each reconnection event drives the system toward complete energy release and a common post-event state.}
  \label{energy_delay_cartoon}
  \end{figure*}

A widely accepted interpretation of the energy-delay correlation is based on a simple conceptual model where magnetic stress gradually builds up over time as the magnetic field becomes increasingly twisted and tangled by continuous footpoint motions, eventually reaching a critical threshold. This threshold can be defined in terms of a critical magnetic stress \citep{Parker_1988, Lopez_2015ApJ...799..128L}. Once this threshold is exceeded, the magnetic field undergoes reconnection, releasing a portion of the available free magnetic energy \citep{Schonfeld_2020ApJ...905..115S}. This process is illustrated in panel (a) of Figure \ref{energy_delay_cartoon}, where the black dashed line denotes the upper threshold or `ceiling' corresponding to the critical stress. In this scenario, larger the energy release, longer it takes for the field to build up stress again and reach the critical state required for another reconnection event.

An alternative picture considers that the magnetic stress relaxes impulsively to a fixed minimum energy state -- releasing all the stored magnetic energy -- following each reconnection event \citep{Cargill_2014ApJ...784...49C}. Consequently, longer the delay between events, more energy is accumulated, and larger the subsequent energy release. This model implies the existence of a lower bound or `floor' in the energy evolution, as depicted by the black dashed line in panel (b) of Figure \ref{energy_delay_cartoon}, representing the minimum energy state the system tends to attain after each event.

Therefore, analysis of the energy-delay correlation of nanoflares provides a valuable diagnostic for constraining the coronal heating models. It enables investigation of key physical questions, such as whether nanoflares occur as independent, randomly distributed events, or if the system exhibits temporal memory indicative of stress accumulation. Additionally, it can help determine whether a longer quiescent period leads to stronger energy release, reflecting underlying energy storage and release dynamics in the coronal magnetic field.

Hydrodynamic simulations, or `loop models', have been widely used to study nanoflare heating by comparing synthetic and observed emission measure distributions. These models typically vary the temporal and energetic properties of nanoflares to reproduce observations. Hence the input parameters are often not physically constrained. In particular, the time delay between successive nanoflares has always been an ad hoc input to these models. For example, \citet{Reep_2013ApJ...764..193R} adopted a constant delay, while subsequent studies assumed delays that scale with the energy of the next event \citep{Cargill_2014ApJ...784...49C, Bradshaw_2016ApJ...821...63B, Barnes_2019ApJ...885..108B}. Alternatively, \citet{Schonfeld_2020ApJ...905..115S, Mondal_2025ApJ...980...75M} related the delay to the energy of the previous event. 

In a recent study, \cite{Knizhnik_2020SoPh..295...21K} used a magnetically driven 3D MHD model where they tracked discontinuous changes in field line connectivity and found that nanoflare delays follow a power-law distribution with a slope close to $-1$. However, while their results suggest power-law behavior in both delay and energy, the relationship between the two was not examined. This remains an open question and is key to determining whether nanoflare energy release is dictated by the local evolution of magnetic stresses or by interactions across multiple field lines.

In this study, we investigate the correlation between nanoflare energies and delays in a 3D MHD simulation of a subset of an active region. The simulation, described in \cite{Johnston_2025ApJ...994..139J}, self-consistently generates nanoflares from prescribed photospheric driving. Building on this, we performed a field-line-based analysis to study the energy and frequency distributions of the nanoflares across a large ensemble of magnetic field lines \citep{Shanwlee_2025}. For each field line, nanoflare energies and durations were quantified using three different approaches. Here, we examine whether any statistical relationship holds between the energies of these events and the delays between their occurrences. 

The rest of the paper is organized as follows. Section \ref{sec:energy_delay_stats} provides a brief overview of the datasets used in this study. Section \ref{sec:correlation_tests} describes the correlation tests and methodologies employed to assess the correlation between nanoflare energies and delays. Section \ref{sec:Results} presents the results obtained from these analyses. Finally, Section \ref{sec:Discussion} discusses the main findings of this study.

\section{Dataset}\label{sec:energy_delay_stats}
In \cite{Johnston_2025ApJ...994..139J}, we performed a 3D multi-strand simulation of a subset of an active region. The simulation begins with a straight and uniform magnetic field in a computational domain which span $33.75 \times 33.75$ Mm$^2$ in the horizontal  ($x-y$) directions and $130$ Mm in the vertical ($z$) direction. Motions representing photospheric convection are applied at the two ends of the photospheric boundaries in the vertical direction. These motions twist and tangle the field lines, forming thin current sheets at the interfaces between quasi-independent magnetic strands. These current sheets eventually dissipate, releasing the stored magnetic energy through small, impulsive heating events. We tracked the evolution of 3136 magnetic footpoints from the bottom driving plane, following the prescribed flows and tracing the field lines upward to the top plane \citep{Shanwlee_2025}. Nanoflares occurring on the field lines were identified using three distinct methods -- named as Method A, B and C -- each with its own strengths and limitations. 

A comprehensive description of the  methods can be found in \cite{Shanwlee_2025}. Briefly, Methods A and B identify magnetic reconnection events based on abrupt jumps in the connectivity between the lower and upper photospheric boundaries. Once identified, energies were measured by integrating the viscous shock heating along the coronal portion of the field line using either a fixed nanoflare duration (Method A) or taking the duration to be the full width at half maximum (FWHM) of the temporal heating profile (Method B). Method C does not rely on connectivity changes; instead, it identifies events directly from the heating profile using a peak-finding approach, with event duration defined as the full width at half maximum measured from the peak above the baseline.

Unlike the conventional definition of a nanoflare--which involves the entire bundle of magnetic flux that experiences impulsive heating–we define nanoflares as heating events experienced by individual field lines. In the traditional view, nanoflare energy scales with the area occupied by the reconnecting flux bundle. In contrast, our definition assigns nanoflare energy to the total energy gained by a single field line during an event. Note that an event can involve multiple connectivity jumps. 3D reconnection occurs continuously over a finite duration as the field line passes through the current sheet diffusion region. This is known as flipping reconnection \citep{Pontin_2005PhPl...12e2307P} and it appears as string of reconnection jumps in our simulation output. Occasionally, a field line may reconnect at two distinct reconnection sites at nearly the same time, in which case we treat them as one nanoflare. It is considerably more common for a field line reconnecting at a single site to exhibit multiple heating peaks. The second nanoflare from Method B (green) in Figure 8 of \cite{Shanwlee_2025} is a good example. Secondary peaks could be directly related to the single-site reconnection, or, for example, they could be due to compressive heating associated with the deceleration of a reconnection jet from a separate nearby site. Note that the total duration of the nanoflare is quite short ($\sim$ 100 s) and much shorter than the plasma cooling time. Note also that many of the numerous nanoflares identified by method C do not have connectivity jumps that exceed our threshold for reconnection. Nanoflare energies, in our simulation, are computed by spatially integrating the viscous heating rate over the coronal segment of the field line and temporally integrating over the duration of the event. As a result, nanoflare energies are expressed per field line or per unit area.

Of course, field lines are infinitely thin. As used here in the context of nanoflare energy, a ``field line'' is an extremely thin flux tube. Its cross-sectional area is essentially constant because the plasma beta is small and the field has minimal curvature. The initially straight and uniform guide field dominates at all times, thereby maintaining a nearly constant magnetic pressure over the entire coronal domain. The shape of the cross section may vary along the flux tube even though its area is constant. This has no influence on the field-aligned physics that determine the thermal properties of the plasma.

For any given nanoflare, we define the energy of the subsequent event as $E_{next}$, and the energy of the current event as $E_{previous}$. Using the known start and end times of each nanoflare, we estimate the inter-event delay ($\tau_D$) as :
\begin{equation}
    \tau_D = \tau_{start}^{next} - \tau_{end}^{previous}
\end{equation}
where, $\tau_{\rm start}^{\rm next}$ denotes the start time of the event with energy $E_{\rm next}$, and $\tau_{\rm end}^{\rm previous}$ denotes the end time of the event with energy $E_{\rm previous}$. 

\section{Tests for correlation}\label{sec:correlation_tests}
For our analysis, we have chosen to use the weighted $t-$ statistic ($t_w$) described in \cite{Efron_1992ApJ...399..345E} and used by \cite{Lee_1993ApJ...412..401L, Porter_1995ApJ...454..499P, Klimchuk_2020ApJ...900..167K} and the Spearman rank correlation \citep{Spearman_ca468a70-0be4-389a-b0b9-5dd1ff52b33f}. Both are nonparametric statistical methods that assess correlations based on the rank ordering of data rather than their values. Specifically, a nonparametric correlation test of data set $(x_i, y_i)$ evaluates whether the largest value of $y$ corresponds to largest values of $x$, the second largest value of $y$ corresponds to the second largest value of $x$, and so on, independent of the actual magnitudes. This rank-based approach makes these methods robust to outliers. Additionally, they require no assumptions about the underlying distribution of the data, unlike standard correlation tests, which assume measured values that are normally distributed about the true value.

Our statistical analysis includes two experiments. \textit{First}, we conduct a null hypothesis test which evaluates the probability of uncorrelation between the energy and delay datasets. \textit{Second}, we assume a power-law relationship  \footnote[1]{We adopt a power-law dependence between $E$ and $\tau_D$ because many physical systems obey power laws. Other forms are possible. We note that scenario b in Figure \ref{energy_delay_cartoon} would predict a quadratic dependence ($\alpha$ = 2). The energy buildup in the driven magnetic field is the time integral of the Poynting flux, and the Poynting flux increases linearly with time because the stress component of the field increases linearly with time. Hence, the energy builds up as $t^2$.} of the form $E \propto \tau_D^{\alpha}$ and estimate the value of $\alpha$. Under this assumption, we determine: (a) the direction and strength of the {\it dependence}, indicated by the sign and magnitude of $\alpha$, where $\alpha > 0$ implies a direct correlation and $\alpha < 0$ indicates an inverse or anti-correlation; and (b) the strength of the {\it correlation}, which reflects how tightly the $(\tau_D, E)$ data points align along a straight line in log-log space.

To estimate the value of $\alpha$, we use the following logic: if a power-law relationship $E \propto \tau_D^\alpha$ exists, then for the correct value of $\alpha$, the ratio $E/\tau_D^\alpha$ should be constant. This transformation should eliminate any correlation between $\tau_D$ and $E/\tau_D^\alpha$. Therefore, the most probable value of $\alpha$ is the one that maximizes the probability of uncorrelation between $\tau_D$ and $E/\tau_D^\alpha$. A detailed description of this method can be found in \cite{Porter_1995ApJ...454..499P}. 

\begin{figure*}
 \centering
 \includegraphics[width=0.95\textwidth,angle=0]{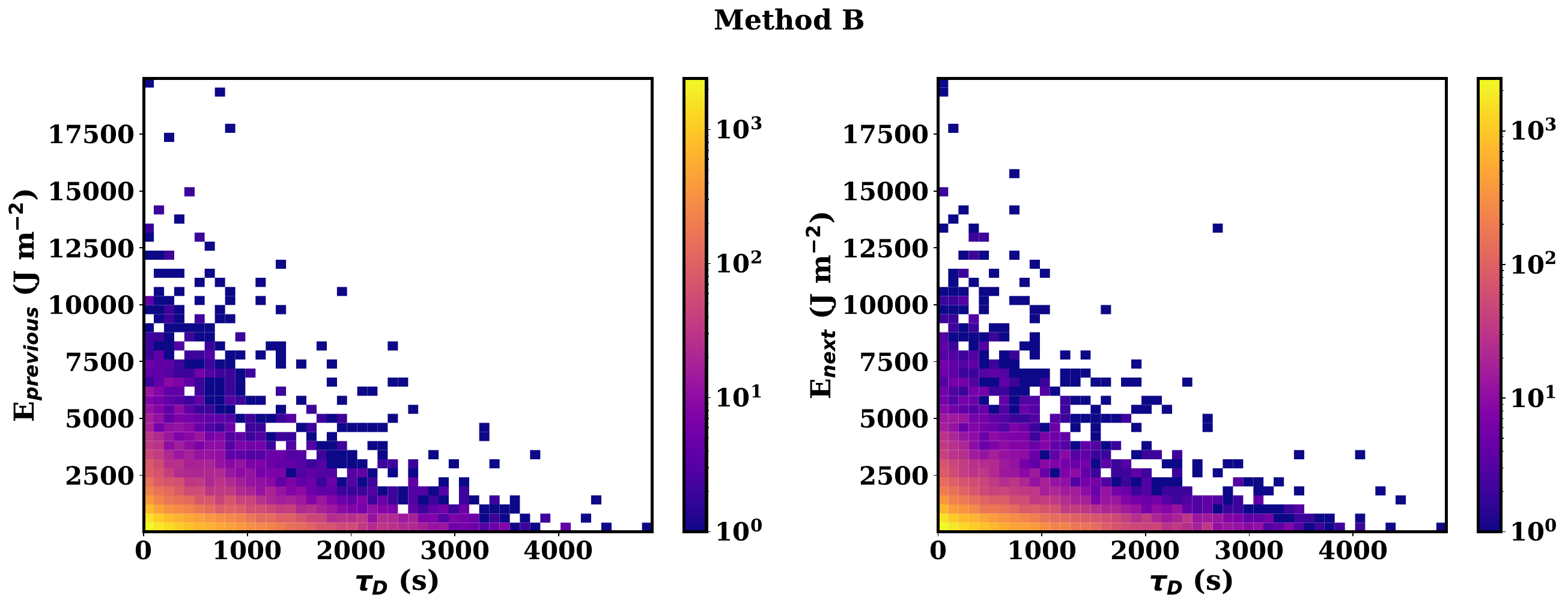}
\includegraphics[width=0.95\textwidth,angle=0]{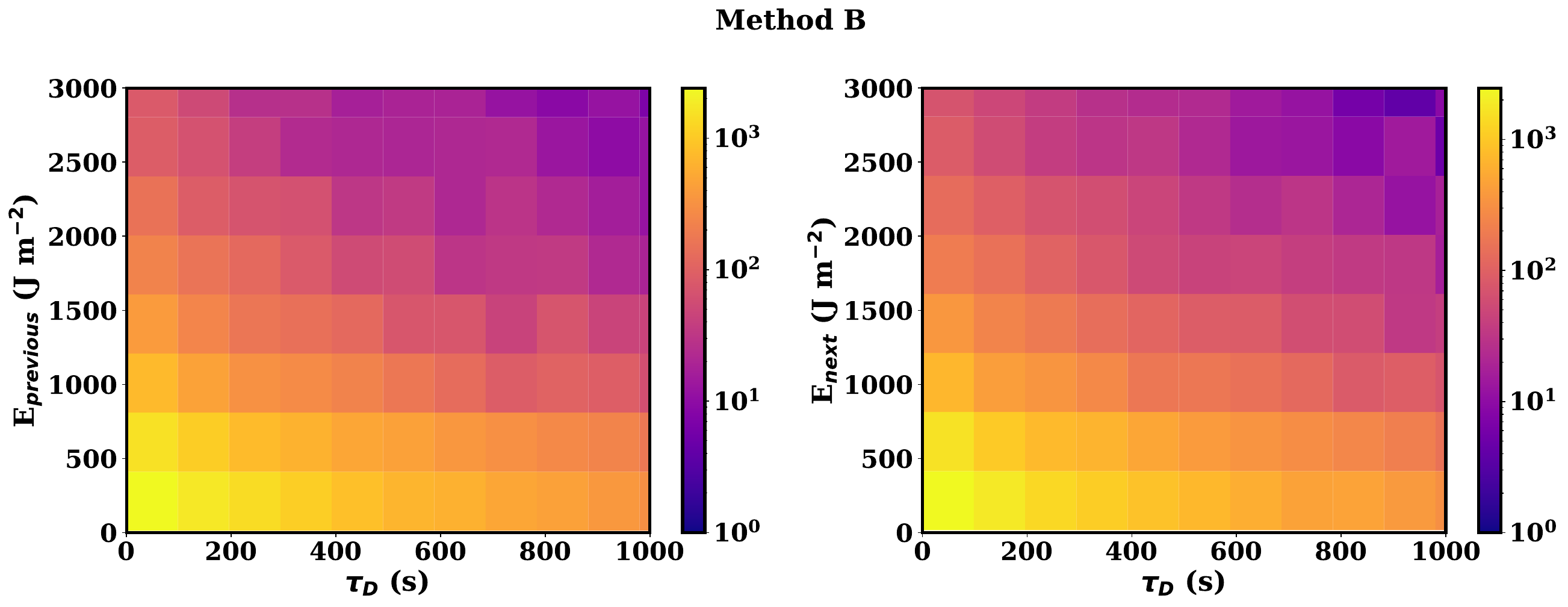}
 \caption{2D histograms of nanoflare energies and delays for events identified using Method B. The left (right) panels correspond to energies from the previous (next) event, respectively. The lower panels provide close-ups of the upper panels.}
  \label{2D_hist_MethB}
  \end{figure*}

\section{Results}\label{sec:Results}
We begin by sampling a subset from the original $(\tau_D, E)$ data set. This is done by randomly selecting 1000 unique values of one variable and retrieving the corresponding values of the other, ensuring that all entries are unique to avoid rank ties. We have also tested with 10,000 samples and found consistent results. For each subset, we perform the following analyses:

(a) \textit{Probability of uncorrelation (p-value):} We perform the $t_w$ and Spearman rank correlation tests to assess the degree of correlation between $\tau_D$ and $E$. The $t_w$-test calculates a statistical value (using the ranks of the two variables) and an associated p-value indicating the probability of obtaining the statistical value (or higher) if there exists no correlation between $\tau_D$ and $E$ in the chosen subset. A low probability suggests that the variables are likely to be correlated. Similarly, the Spearman test provides the rank correlation coefficient along with its p-value.  

(b) \textit{Most probable value of $\alpha$:} We then estimate the most probable value of $\alpha$ in the scaling relation $E \propto \tau_D^\alpha$ by performing correlation tests on the transformed dataset $(\tau_D, E/\tau_D^\alpha)$ over a chosen range of $\alpha$. For each $\alpha$, we compute the probability of uncorrelation between $\tau_D$ and $E/\tau_D^\alpha$. In the chosen $\alpha$ range, the most probable value of $\alpha$ is identified as the one that maximizes this probability of uncorrelation.

Steps (a) and (b) are repeated over 1000 independent re-samplings to construct statistical distributions (histograms) of the probability of uncorrelation (p-value) and most probable value of $\alpha$. This approach increases the likelihood of detecting any significant correlation that may be present within the original data set. In the following sections, we present the correlation results for nanoflares identified using Method B. The corresponding results from Methods A and C are provided in Appendix \ref{appendix:Method A & C}.

\begin{figure*}
 \centering
 \includegraphics[width=0.95\textwidth,angle=0,page=3]{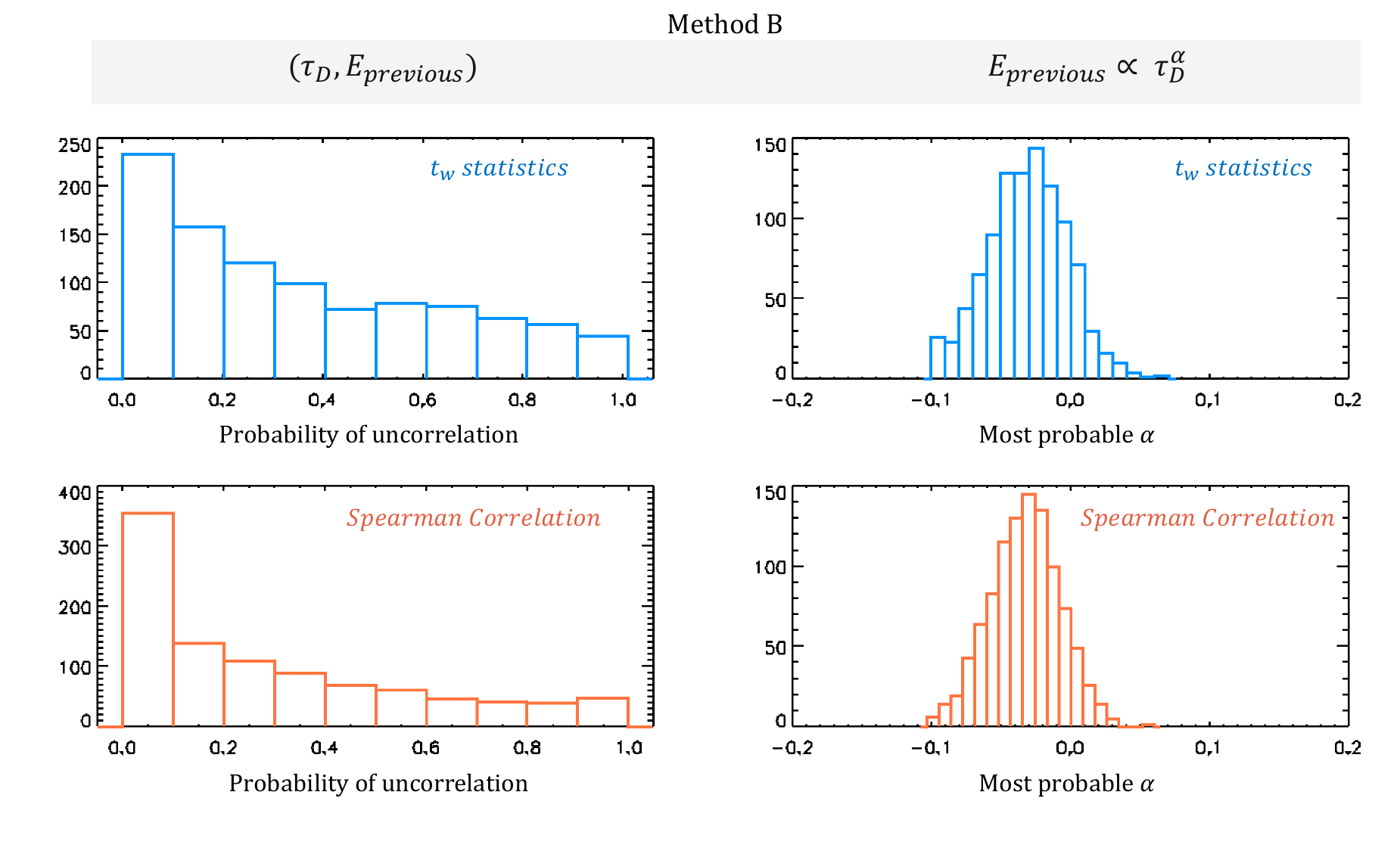}
 \includegraphics[width=0.95\textwidth,angle=0,page=4]{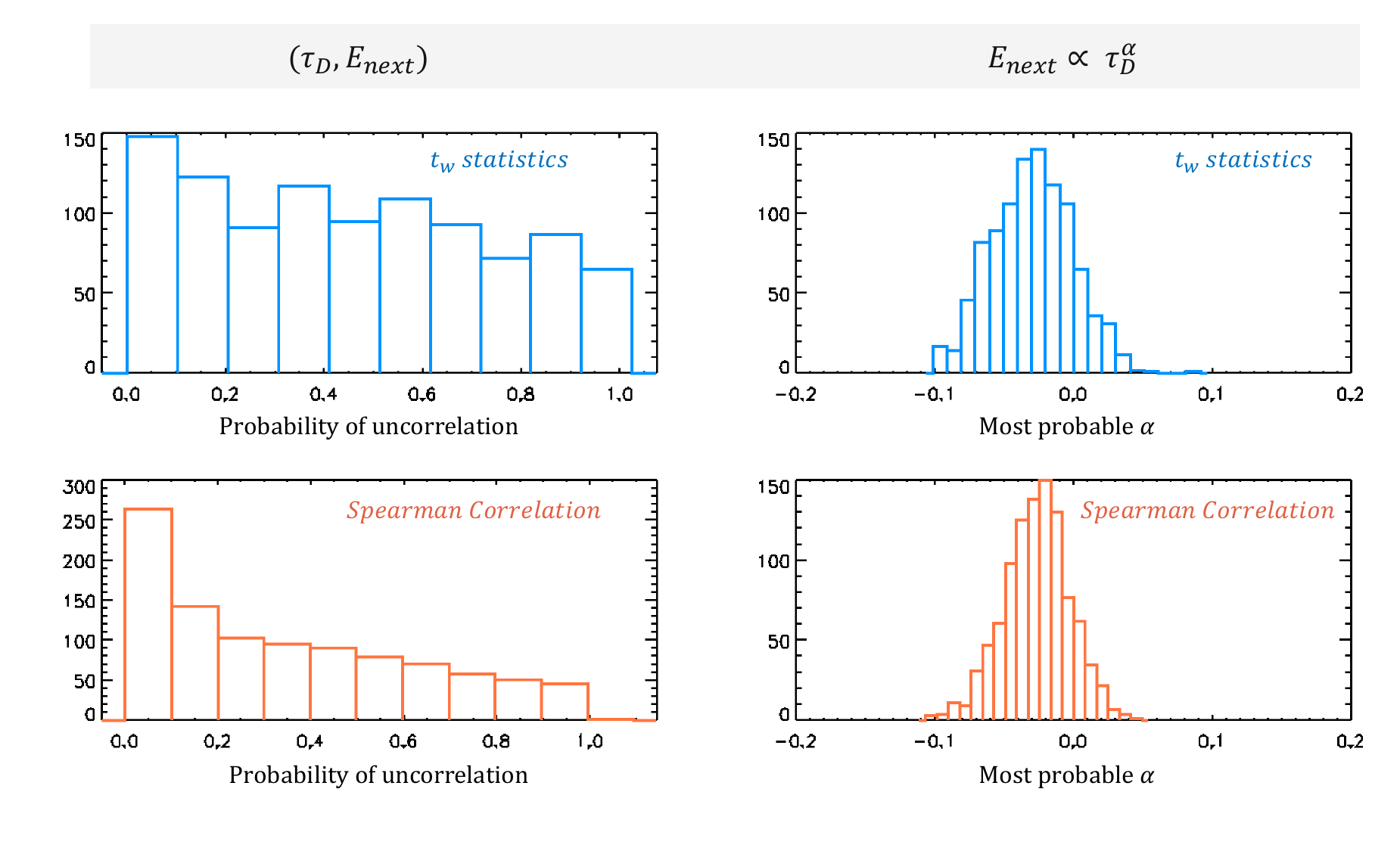}
 \caption{Statistical correlation results obtained via bootstrapping of the original nanoflare sample identified using Method B. The first four histograms correspond to the $(\tau_D, E_{previous})$ data set, and the later four to $(\tau_D, E_{next})$. Blue and red histograms represent results from the $t_w$ and Spearman rank correlation tests, respectively. Left panels show the histograms of the probability of uncorrelation (p-values), while right panels display the most probable values of $\alpha$. A substantial fraction of the sampled subsets suggest the presence of weak or no statistical correlation. The extremely low values of $\alpha$ further support the absence of a strong statistical relationship between $\tau_D$ and either $E_{\text{previous}}$ or $E_{\text{next}}$.}
  \label{MethB_stat}
  \end{figure*}

\subsection{Nanoflares from Method B}
Figure \ref{2D_hist_MethB} presents 2D histograms of nanoflare energies and associated delays as identified by Method B. In the left panel, energies correspond to those of the previous event, whereas in the right panel, they correspond to the next event. The bottom panel provides close-up views of the histogram regions where most events are concentrated, as indicated in the top panel. This zoomed-in representation enables a more detailed examination of the distribution and reveals any subtle variations—if present—that may not be easily visible in the full-scale view. The delay, as mentioned before, is defined as the time interval between the end of one event and the start of the next. Notably, these delays represent the temporal separation between consecutive events, regardless of event energy.

Figure \ref{MethB_stat} shows histograms of the probability of uncorrelation (left panel) and the most probable value of $\alpha$ (right panel), obtained by resampling the original dataset 1000 times \footnote[2]{The correlation code employed for the bootstrapping analysis and the generation of histograms is openly accessible via our Zenodo repository: https://doi.org/10.5281/zenodo.17834438}. The first four histograms correspond to the correlation analysis for the data set ($\tau_D$, $E_{previous}$), while the subsequent four depict the correlation results for ($\tau_D$, $E_{next}$). The blue and red histograms represent the results obtained from the $t_w$ and the Spearman rank correlation tests, respectively.

\subsubsection{Correlation between $\tau_D$ \& $E_{previous}$}
The histograms of the uncorrelation probability between ($\tau_D$, $E_{previous}$) show a peak at low values($<10\%$), suggesting that many subsets exhibit a significant correlation. However, a considerable fraction of the distribution also corresponds to high uncorrelation probabilities, indicating variability across the sampled subsets. While this suggests a broad potential for correlation, the fact that the most probable values of $\alpha$ are consistently close to zero implies that the strength of any such correlation is extremely weak or negligible. 

\subsubsection{Correlation between $\tau_D$ \& $E_{next}$}
The histogram of the probability of uncorrelation obtained from $t_w$ test indicate number of cases with variable probability (low to high) of uncorrelation between $\tau_D$ and $E_{next}$. On the other hand, the Spearman correlation tests performed over the bootstrapped samples, reveal a considerable number of cases with low uncorrelation probability, indicating a general tendency toward correlation. However, a substantial portion of the distribution starts showing high uncorrelation probabilities, suggesting notable variability across subsets. Despite this apparent potential for correlation, the fact that the most probable values of $\alpha$ are consistently near zero indicates a very weak or negligible dependence between $\tau_D$ and $E_{next}$. 

\subsubsection{Spread in the data}
\begin{figure*}
 \centering
 \includegraphics[width=0.95\textwidth,angle=0,page=2]{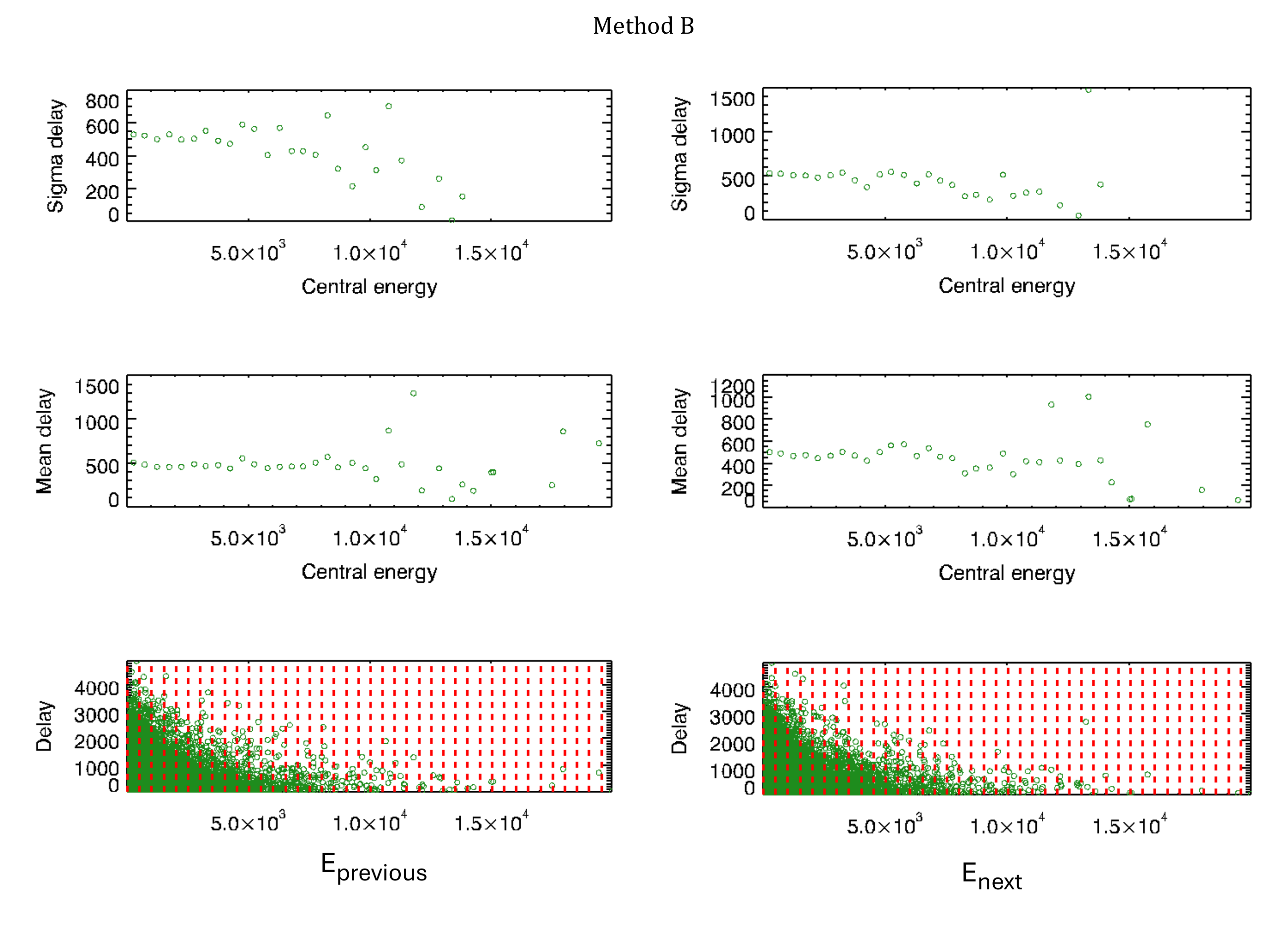}
 \caption{The figure illustrates the spread in nanoflare delays in the original dataset obtained from Method B. The bottom panel shows scatter plots of nanoflare energies versus delays. The middle and top panels display the mean and standard deviation of delays within each energy bin, with the bins defined by the red dashed lines in the bottom panel. The fact that the standard deviations (sigma delays) are comparable to the mean delays in each bin suggests that any correlation between $\tau_D$ and either $E_{previous}$ or $E_{next}$ is likely to be weak or statistically insignificant.}
  \label{Spread_MethB}
  \end{figure*}

Figure \ref{Spread_MethB} illustrates the distribution of delays within selected energy intervals. This is obtained by dividing the entire energy range into uniformly sized bins. The boundaries of these energy bins are indicated by red dashed lines in the bottom panel. For each energy bin, the mean and standard deviation of the corresponding delay values are computed and plotted as a function of the central energy of the bin. The resulting mean and standard deviation of the delays are presented in the middle and top panels, respectively. The plots in the left panel correspond to the analysis of ($\tau_D$, $E_{previous}$), whereas the right panel displays the results derived from the ($\tau_D$, $E_{next}$). It is to be noted that, within each energy bin, the standard deviation of the delays is comparable to the corresponding mean delay. This implies that, for any given energy range, nanoflares recur with delays spanning over a broad range. This strongly confirms the lack of correlation between the measured nanoflare energies and their delays. 

As mentioned before, these results are applicable to nanoflares identified using Method B. The results for the other two methods (A \& C) are very similar. The probability that $\tau_D$ is correlated with either $E_{previous}$ or $E_{next}$ is generally low, though there is evidence of significant correlation for some of the samples. In all cases the dependence of $\tau_D$ on $E_{previous}$ and $E_{next}$ is very weak, with $\alpha$ values close to 0. Finally, the spread of delays in different energy bins is large, with the standard deviation comparable to the mean. We conclude that there is no significant relationship between the nanoflare delays and their corresponding energies (previous or next). The statistical correlation results for Method A \& C are discussed in Appendix \ref{appendix:Method A & C}.

Results discussed above apply to nanoflares regardless of their energies. That is, the results show no correlation between consecutive nanoflares. To investigate whether any correlation exists specifically for high-energy nanoflares, we performed the same statistical tests on high-energy nanoflares. The outcomes remain consistent with the full dataset--no significant correlation is found between nanoflare delays and the energies of either previous or next events. The corresponding histograms supporting this conclusion are provided in the Appendix \ref{appendix:high energy} for high energy nanoflares obtained from all the three methods, separately.

\section{Discussion}\label{sec:Discussion}
The nanoflares investigated in this study are generated self-consistently within a magnetically driven solar active region, as described in \cite{Johnston_2025ApJ...994..139J}. Their energies and durations are subsequently quantified using three distinct methods outlined in \cite{Shanwlee_2025}.
In the present analysis, we examine whether any statistically significant correlation exists between the nanoflare energies and their delays. To identify potential correlations, we applied two non-parametric statistical tests--weighted $t_w$ and Spearman rank correlation--to the nanoflare populations identified using all three methods. Across all methods, the results indicate a generally very weak, if any, correlation between the energies of consecutive nanoflares and their corresponding delays. For some sampled subsets of the data there is a stronger correlation. We also investigated the dependence of energy on delay under the assumption that $E \propto \tau_D^\alpha$. This is strictly only meaningful if there is a significant correlation.  In all cases the most probable value of the exponent $\alpha$ is close to zero, indicating that any dependence is extremely weak. Additionally, the broad distribution of delays within individual energy bins further supports the lack of a strong relationship between nanoflare energy and delay. These conclusions also hold for subsets of high-energy nanoflares and are true whether the delays are compared with the energies of the previous events ($E_{previous}$) or with those of the subsequent events ($E_{next}$).

\begin{figure*}
\centering
\includegraphics[width=0.75\textwidth,angle=0]{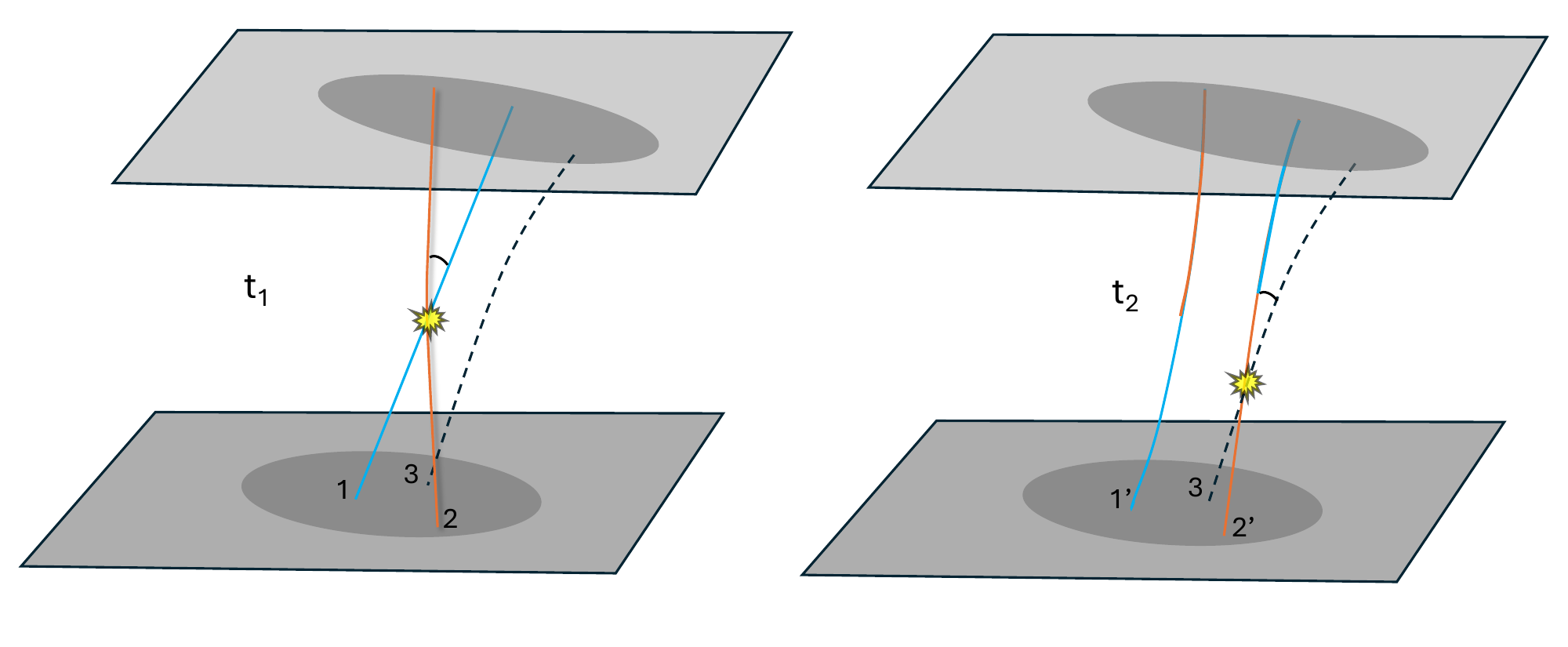}
\caption{Cartoon illustrating how the evolving local magnetic geometry disrupts any correlation between nanoflare energies and their delays.}
\label{Cartoon}
\end{figure*}

Regarding an energy-delay correlation, or lack thereof, we do not expect significant differences between Methods A and B. The main difference between the methods is that nanoflares have longer duration and larger energy in Method B simply because some of the heating profile is excluded in Method A. The delays between nanoflares are very similar in the two cases. Method C differs significantly from the other two in that many of the nanoflares occur on field lines that do not reconnect – at least as identified by our connectivity jump threshold. A decrease in magnetic energy is therefore not expected, as it is for the nanoflares in Methods A and B. The scenarios in Figure \ref{energy_delay_cartoon} may not apply and therefore an energy-delay correlation is less expected.

The lack of relationship between nanoflare energies and delays can partly be explained by the complex magnetic structure of the solar corona. 
The corona can be imagined as a collection of quasi-independent magnetic strands. Reconnection occurs at the current sheet boundaries between the strands. Footpoint driving causes the misalignment between adjacent stands (shear) and associated magnetic free energy to slowly increase. At the same time, the current sheet thickness slowly decreases from the buildup of magnetic pressure. Reconnection occurs when either the thickness  \citep{Leake_2020ApJ...891...62L, Leake_2024ApJ...973...21L} or the shear \citep{Klimchuk_2023FrP....1198194K} reaches a critical value. In the simple binary picture of two strands, we expect a relationship between the amount of energy released and the energy buildup time until the next event, i.e., nanoflare delay. This is indicated schematically on the left side of Figure \ref{energy_delay_cartoon}. However, in a realistic simulation such as ours, each strand is in contact with more than one partner. When reconnection occurs between two strands, there is a rapid restructuring of the field along the entire length of the strands. This may bring one of the strands to super-critical conditions with a third strand, which triggers another reconnection event. This breaks any relationship involving a slow buildup time. 

A possible three-strand interaction is indicated schematically in Figure \ref{Cartoon}. Strands 1 and 2 are oriented such that they meet the critical angle required for reconnection. Nearby, a third strand (dashed line) is initially not favorably aligned--its inclination relative to neighbors is less than the critical threshold--so it does not reconnect (Scenario: $t_1$). However, once the first two strands reconnect, the magnetic field in the region is rearranged. This reconfiguration causes the angle between one of the newly reconnected strands and the third strand to become super-critical (Scenario: $t_2$), in which case a second reconnection occurs.

In this example, the third strand reaches super-critical conditions and reconnects well before it would do so from slow photospheric driving. This destroys a simple relationship between nanoflare energy and delay that would otherwise exist. The time delay until the next event is not governed by the amount of energy released in the previous event, but by external changes in the local magnetic topology.

Multi-strand complexity also impacts the idealized scenario depicted on the right side of Figure \ref{energy_delay_cartoon}. Whether two reconnecting strands can reach a `ground energy state' depends on whether they are interlinked, or braided, with other strands. For example, the kink formed when two isolated strands reconnect will disappear as the strands straighten out. If the strands are braided with other strands, then only a partial straightening is possible and therefore a smaller energy release. This destroys any simple relationship that might otherwise exist between the amount of energy released and the delay since the previous event.

An avalanche of reconnection events involving many strands, sometimes called a `nanoflare storm', is a likely explanation of the distinct coronal loops that are seen in coronal images \cite{Klimchuk_2015,Klimchuk_2023ApJ...942...10K, Johnston_2025ApJ...994..139J}. The loops are bundles of multiple unresolved strands. In contrast, the diffuse component of the corona is believed to be caused by random, uncorrelated nanoflares. We plan to examine whether the diffuse component shows stronger evidence of an energy-delay relationship than the bright loop component in the future.

Finally, we note that strand 2 in Figure \ref{Cartoon} experiences two reconnection events in rapid succession. If the time separation is short enough, this would be classified as a single nanoflare, as described in Section 2. However, the measured delays between nanoflares are generally much longer than the measured durations, so this must be an uncommon occurrence. Furthermore, a large majority of nanoflares in our simulation involve a single reconnection site (multiple connectivity jumps are indicative of flipping reconnection rather than multiple distinct reconnection sites).

In summary, we have found based on a highly realistic MHD simulation that there is no significant relationship between the energy of nanoflares and the delay between successive events. We attribute this to the complexity of the magnetic field in which each magnetic strand interacts with other multiple strands. The interaction both modulates the amount of energy that is released by a single reconnection event and often triggers subsequent events.


\section*{Acknowledgements}
SSM, JAK, CDJ, and LKSD were funded by NASA’s Internal Scientist Funding Model (competed work package program) at Goddard Space Flight Center. Simulations were performed on NASA’s High End Computing Facilities by CDJ. Source code, raw data, and plotting routines are available upon request from the authors. We would like to thank James E. Leake, N. Dylan Kee, Yi-Min Huang, and Veronika Jercic for fruitful discussions. 
We sincerely thank the anonymous referee for their constructive comments and valuable suggestions, which have helped improve the quality and clarity of this manuscript.


\appendix

\section{Correlation Results for Nanoflares from Method A \&  C}\label{appendix:Method A & C}
\begin{figure*}
 \centering
 \includegraphics[width=0.95\textwidth,angle=0]{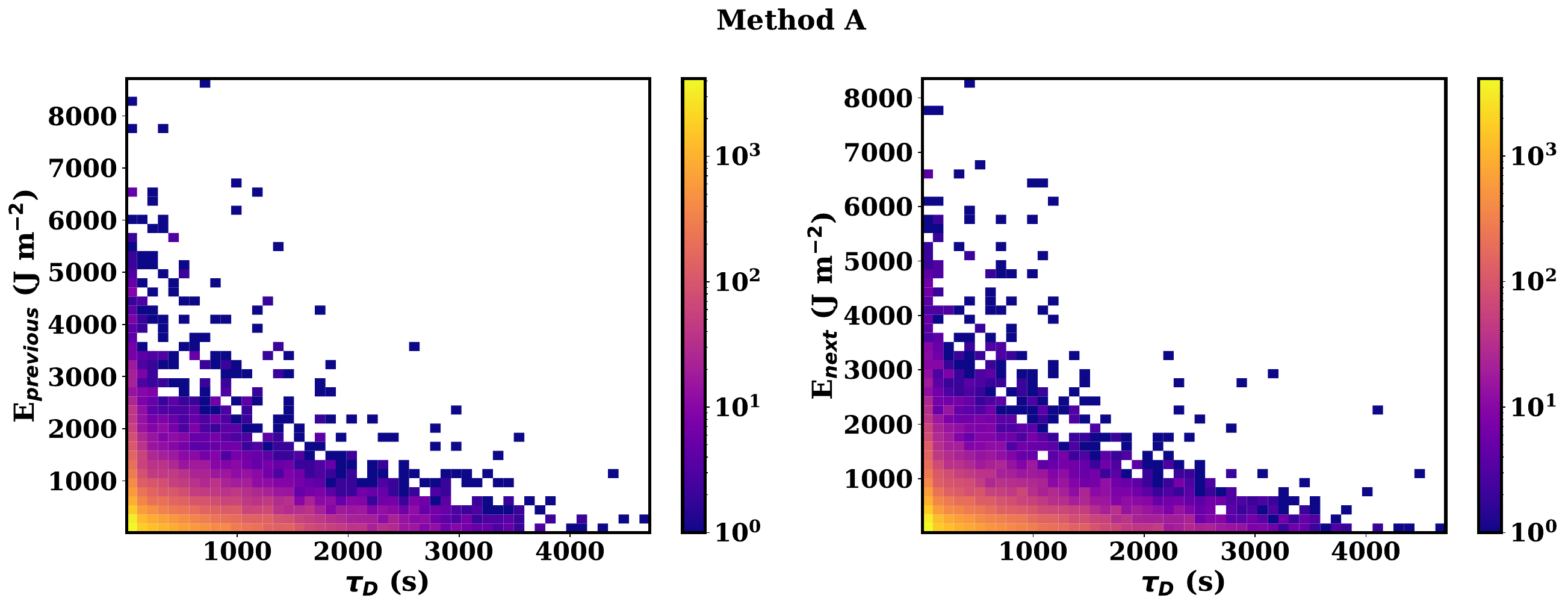}
 \includegraphics[width=0.95\textwidth,angle=0]{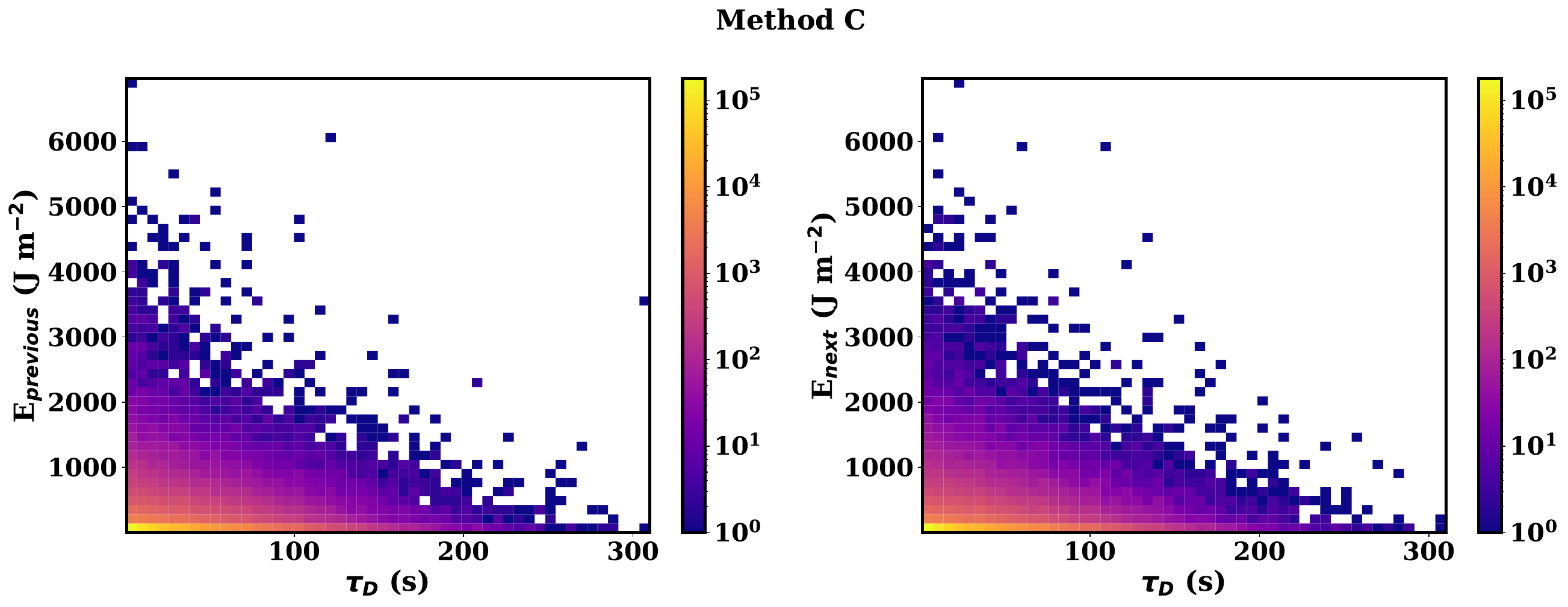}
 \caption{2D histograms of nanoflare energies and delays for events identified using Method A (top panel) and C (bottom panel).}
  \label{2D_hist_MethAC}
  \end{figure*}

\begin{figure*}
 \centering
 \includegraphics[width=0.95\textwidth,angle=0,page=1]{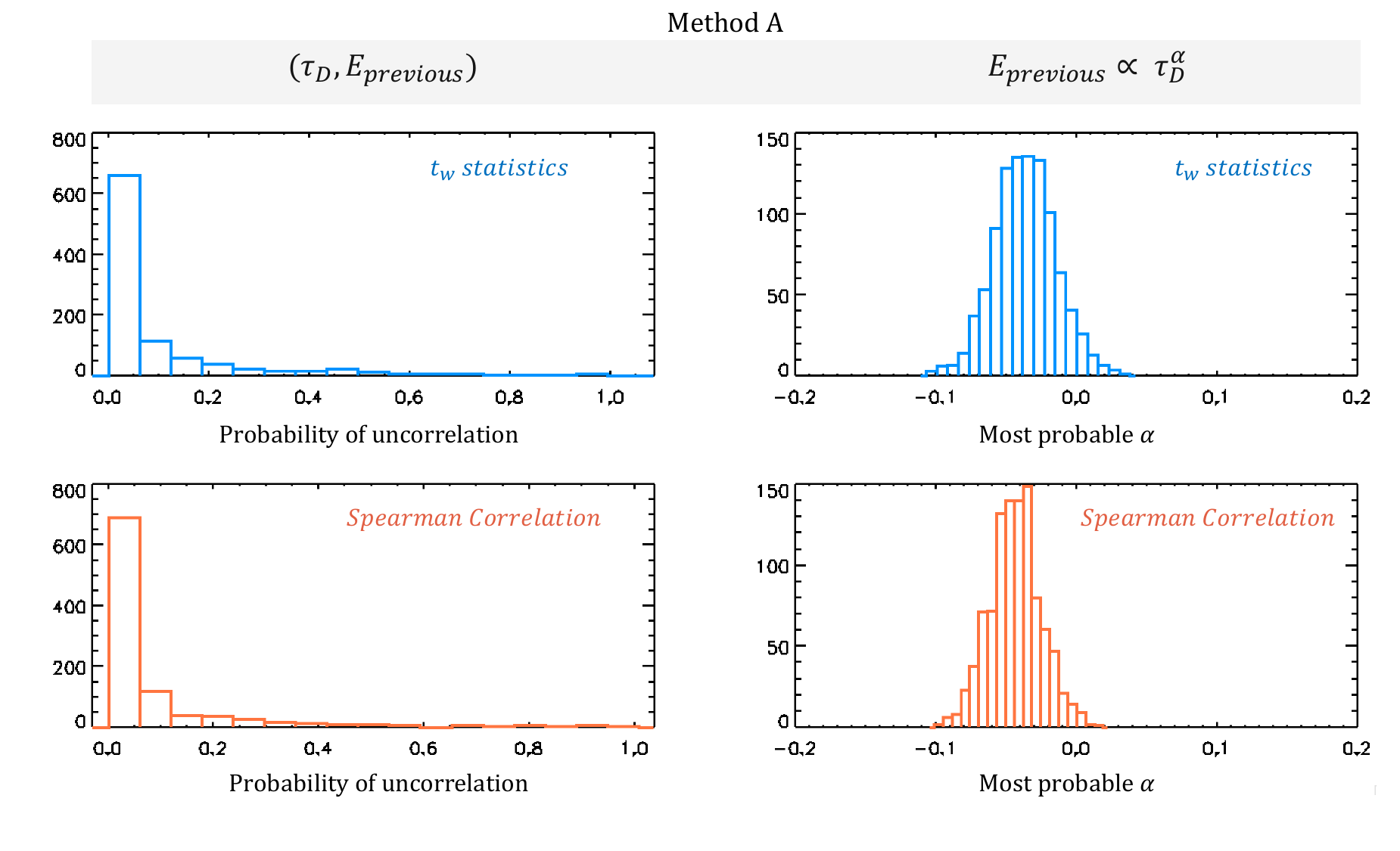}
 \includegraphics[width=0.95\textwidth,angle=0,page=2]{figures/Stat_results.pdf}
 \caption{Statistical correlation results obtained via bootstrapping of the original nanoflare sample identified using Method A. The meaning of each histogram is the same as described in Figure \ref{MethB_stat}. Although the low p-values indicate statistically significant correlations, the clustering of $\alpha$ values near zero suggests no strong or meaningful dependence between $\tau_D$ and either $E_{previous}$ or $E_{next}$.}
  \label{MethA_stat}
  \end{figure*}

\begin{figure}
 \centering
 \includegraphics[width=0.95\textwidth,angle=0,page=5]{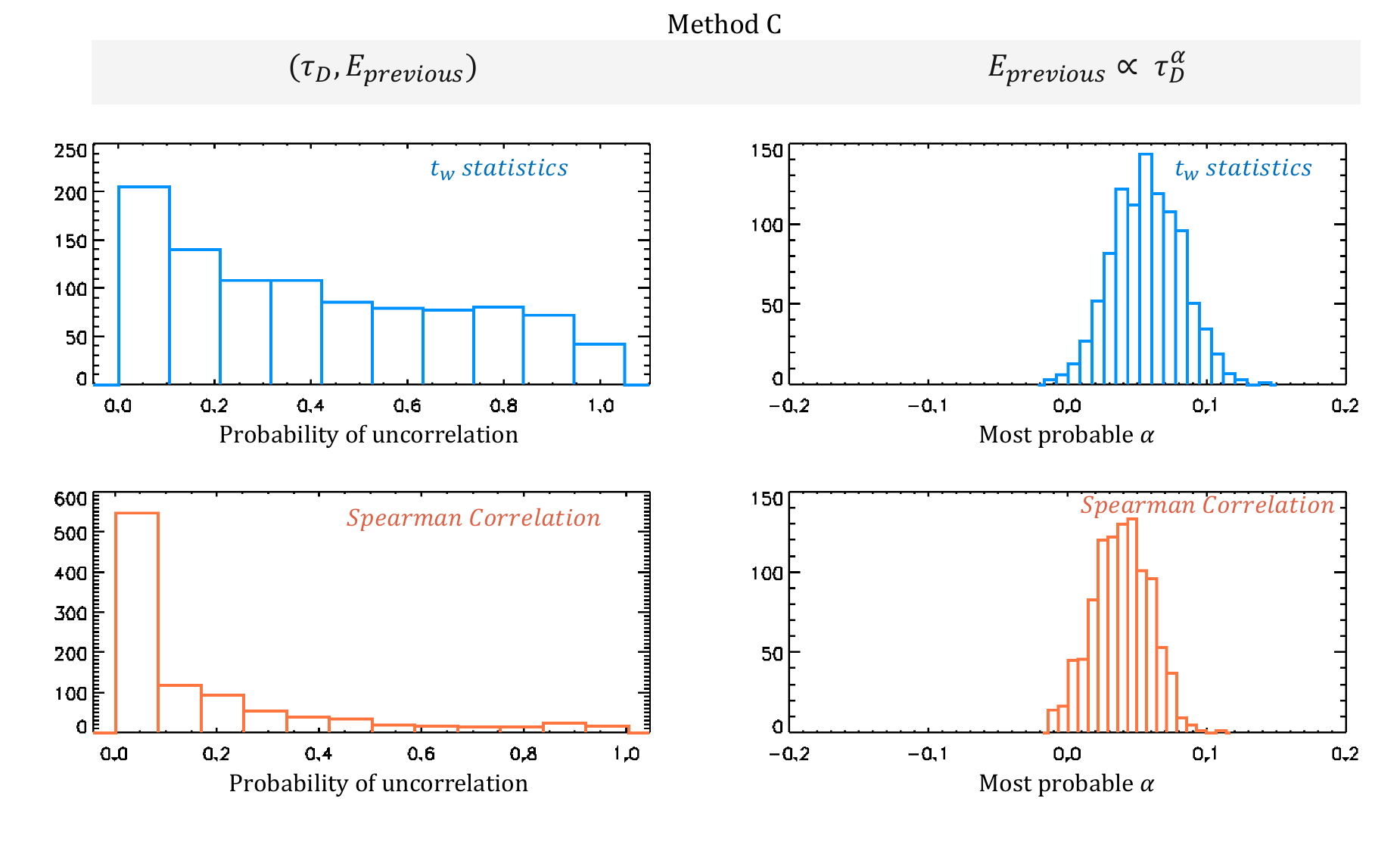}
 \includegraphics[width=0.95\textwidth,angle=0,page=6]{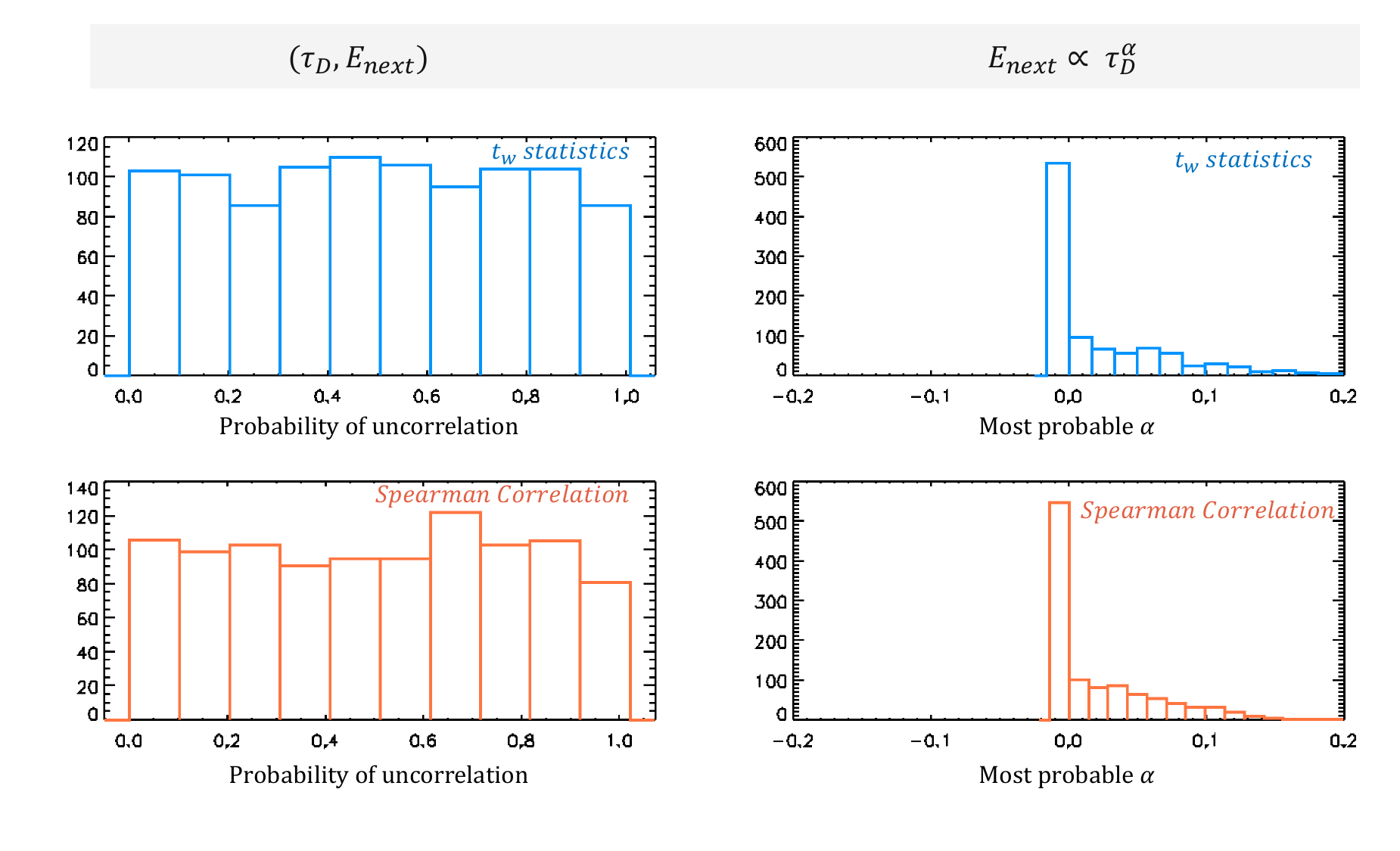}
\caption{Statistical correlation results obtained via bootstrapping of the original nanoflare sample identified using Method C. The meaning of each histogram is the same as described in Figure \ref{MethB_stat}. A substantial fraction of the resampled subsets indicate weak or no correlation, and the extremely small values of $\alpha$ further confirm the absence of a strong statistical relationship between $\tau_D$ and either $E_{\text{previous}}$ or $E_{\text{next}}$.}
\label{MethC_stat}
\end{figure}

Figure \ref{2D_hist_MethAC} shows the 2D histograms of nanoflare energies and associated delays as determined by Method A (top panel) and C (bottom panel), respectively. Figures \ref{MethA_stat} and \ref{MethC_stat} show correlation results obtained for nanoflares identified using Method A and C, respectively. For nanoflares in Method A, the statistical tests suggest a high likelihood of correlation between nanoflare energies and delays, with probabilities often exceeding 90\%. In contrast, Method C shows a substantial fraction of resampled datasets with high probabilities of no correlation, indicating that any correlation is not reliably present across the dataset. Nevertheless, the most probable values of $\alpha$ remain close to zero in both the cases, confirming that the dependence is extremely weak and effectively negligible. Method A yields slightly negative $\alpha$ values and Method C slightly positive ones, but their magnitudes are so small that the sign carries no physical significance. Overall, despite statistical indications, both methods point to the absence of a meaningful correlation, a result consistent for both $E_{previous}$ and $E_{next}$.

\begin{figure}
 \centering
 \includegraphics[width=0.85\textwidth,angle=0,page=1]{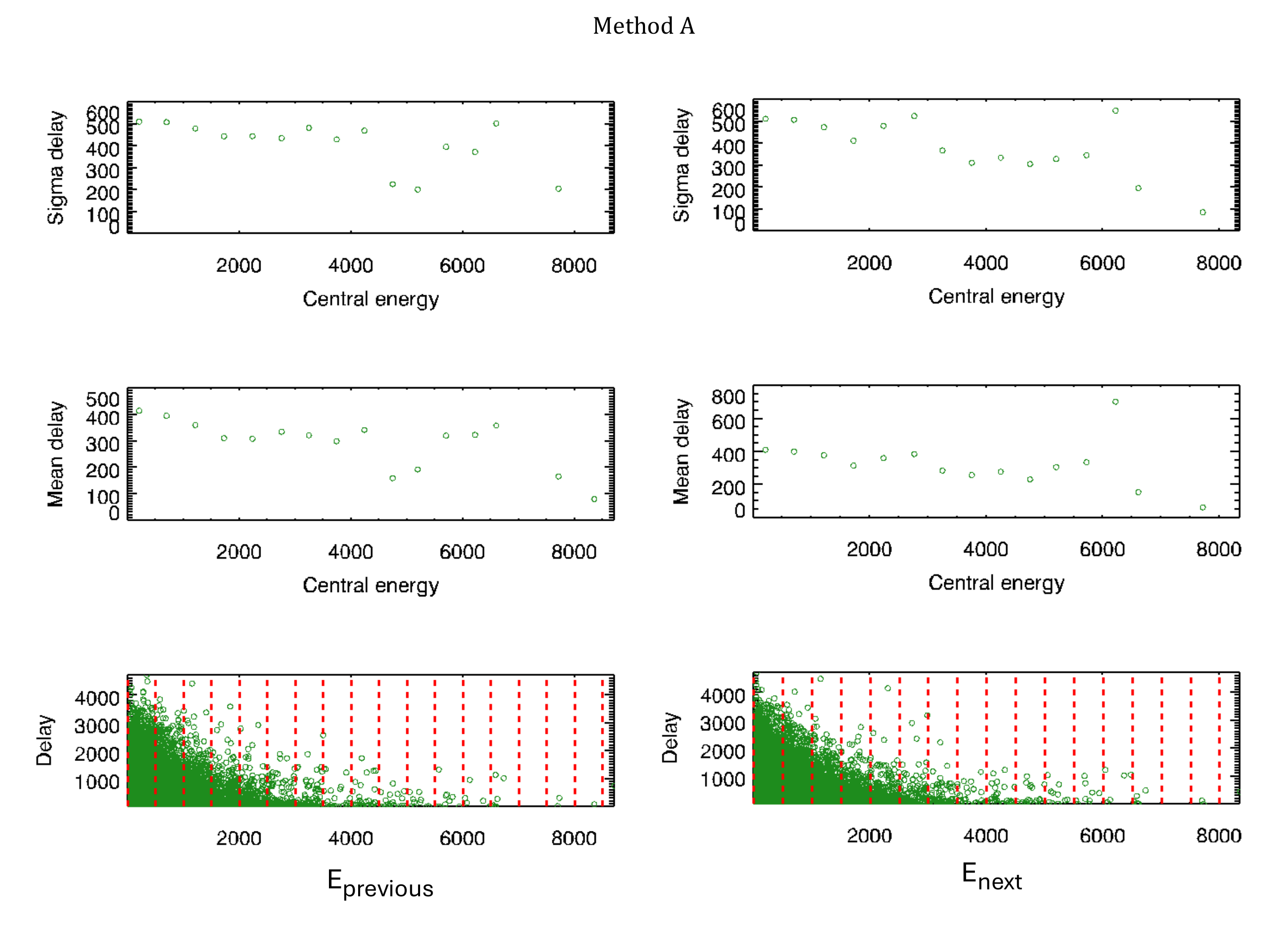}
\includegraphics[width=0.85\textwidth,angle=0,page=3]{figures/Spread_nw.pdf}
 \caption{The figure illustrates the spread in nanoflare delays in the original dataset. First six figures are for nanoflares identified in Method A and next six figures are from Method C. The fact that the standard deviations are comparable to the mean delays in each energy bin suggests that any correlation between $\tau_D$ and either $E_{\text{previous}}$ or $E_{\text{next}}$ is likely to be weak or statistically insignificant.}
  \label{Spread_MethAC}
  \end{figure}

Figure \ref{Spread_MethAC} shows that, for all events, whether detected using Method A (first six plots) or Method C (subsequent six plots), the standard deviation of delays is consistently comparable to the mean delay across all energy bins. As in the previous analysis, the energy range is divided into uniformly sized bins, and the mean and standard deviation of the delays are computed within each bin. The results again reveal a broad spread of delay values within each energy bin, supporting the lack of correlation between nanoflare energies and delays.

\section{Correlation between high-energy nanoflares and their corresponding delays}\label{appendix:high energy}
To investigate whether any correlation is present among high-energy nanoflares, we applied the same statistical tests to a high-energy subset by selecting an energy threshold (different for different methods of nanoflare detection) and analyzing the delays between nanoflares with energies exceeding this threshold. Each threshold is chosen so that the events above it account for one-third of the total energy in the distribution. The results remain consistent with those from the full dataset: no significant correlation is detected between nanoflare delays and the energies of either preceding or subsequent events. Due to the smaller size of the high-energy dataset, each subset now comprises of 100 random selections, with the entire bootstrapping procedure repeated 100 times. Figure \ref{high_energy} presents histograms of the uncorrelation probabilities and the most probable values of $\alpha$ obtained from both the $t_w$ and Spearman rank correlation tests for nanoflares identified using Methods A, B, and C. In all cases, the histograms show that the estimated $\alpha$ values peak near zero, indicating at most an extremely weak dependence between high-energy nanoflares and their delays.

 \begin{figure*}
 \centering
 \includegraphics[width=0.95\textwidth,angle=0,page=1]{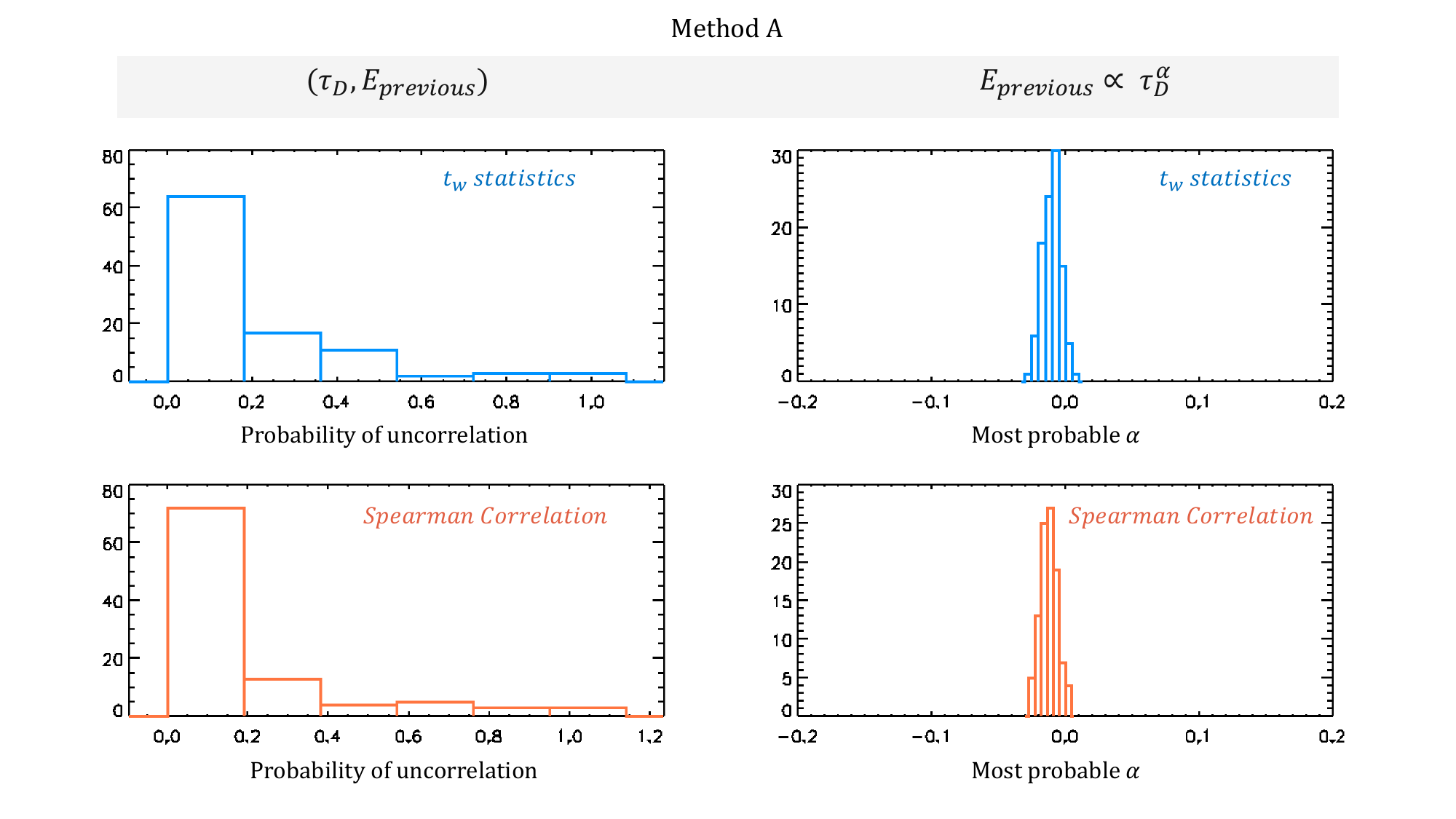}
 \includegraphics[width=0.95\textwidth,angle=0,page=2]{figures/Stat_results_high_energies.pdf}
 \end{figure*}

 \begin{figure*}
 \centering
 \includegraphics[width=0.95\textwidth,angle=0,page=3]{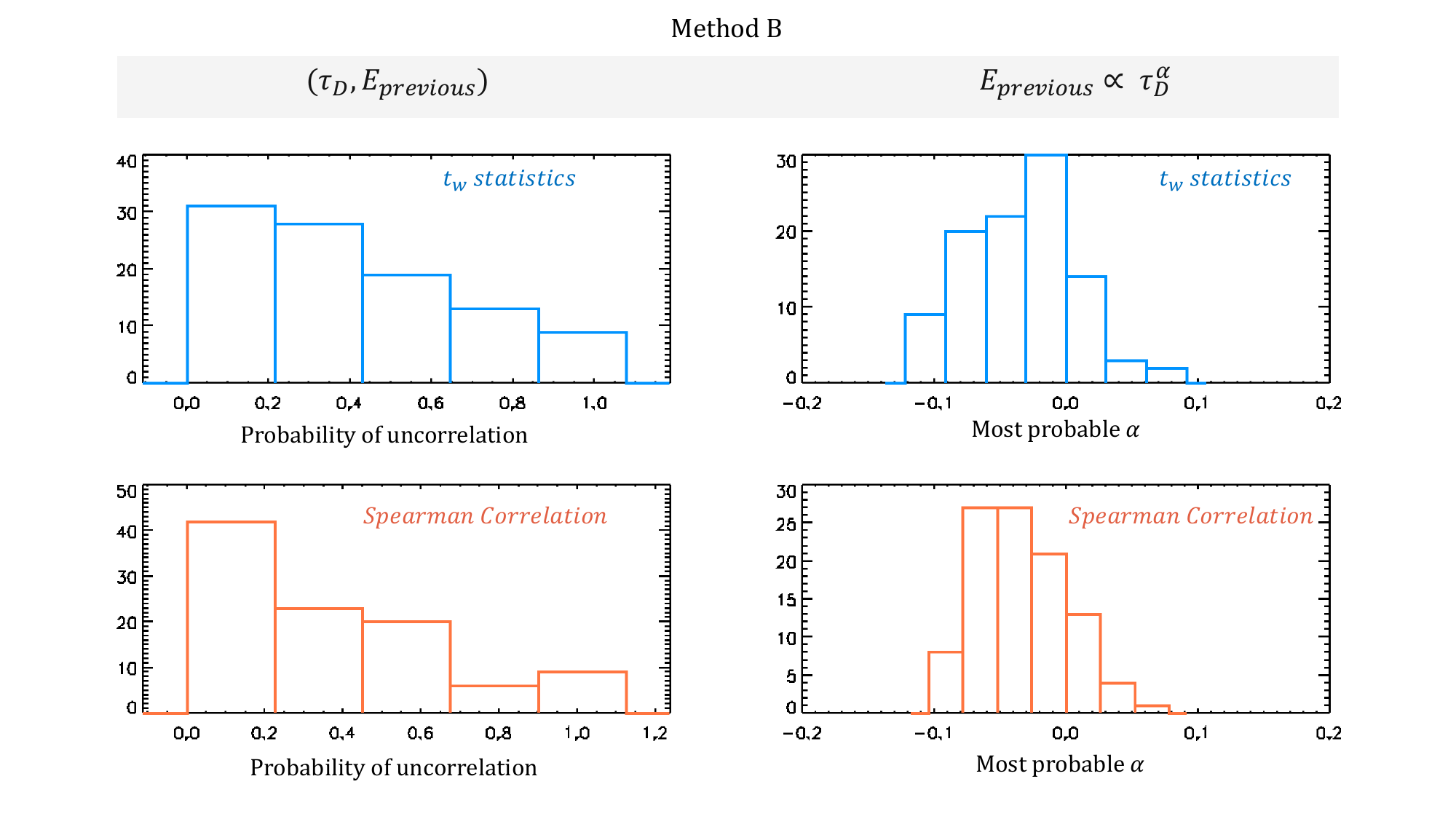}
 \includegraphics[width=0.95\textwidth,angle=0,page=4]{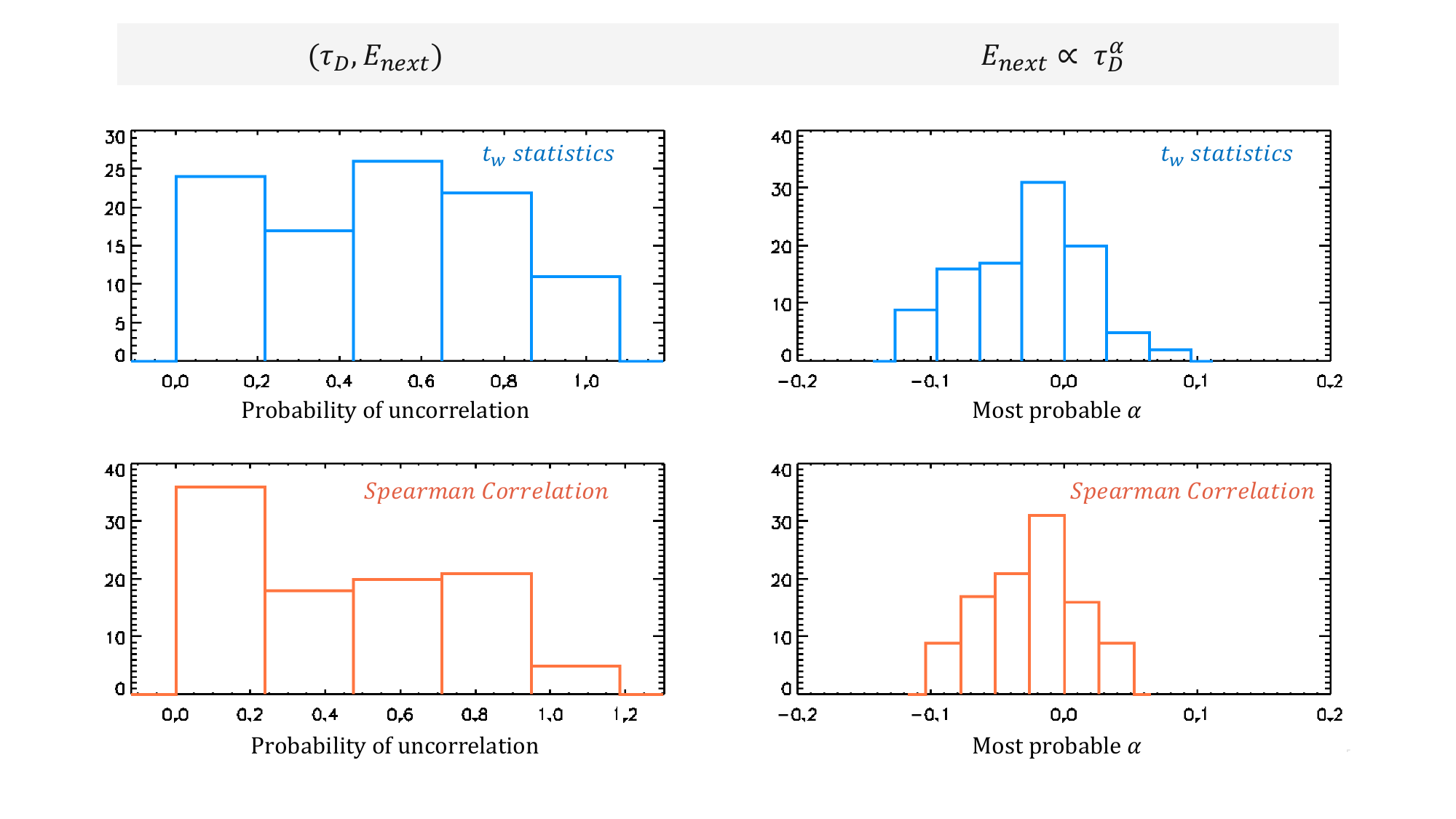}
  \end{figure*}

  \begin{figure*}
 \centering
 \includegraphics[width=0.95\textwidth,angle=0,page=5]{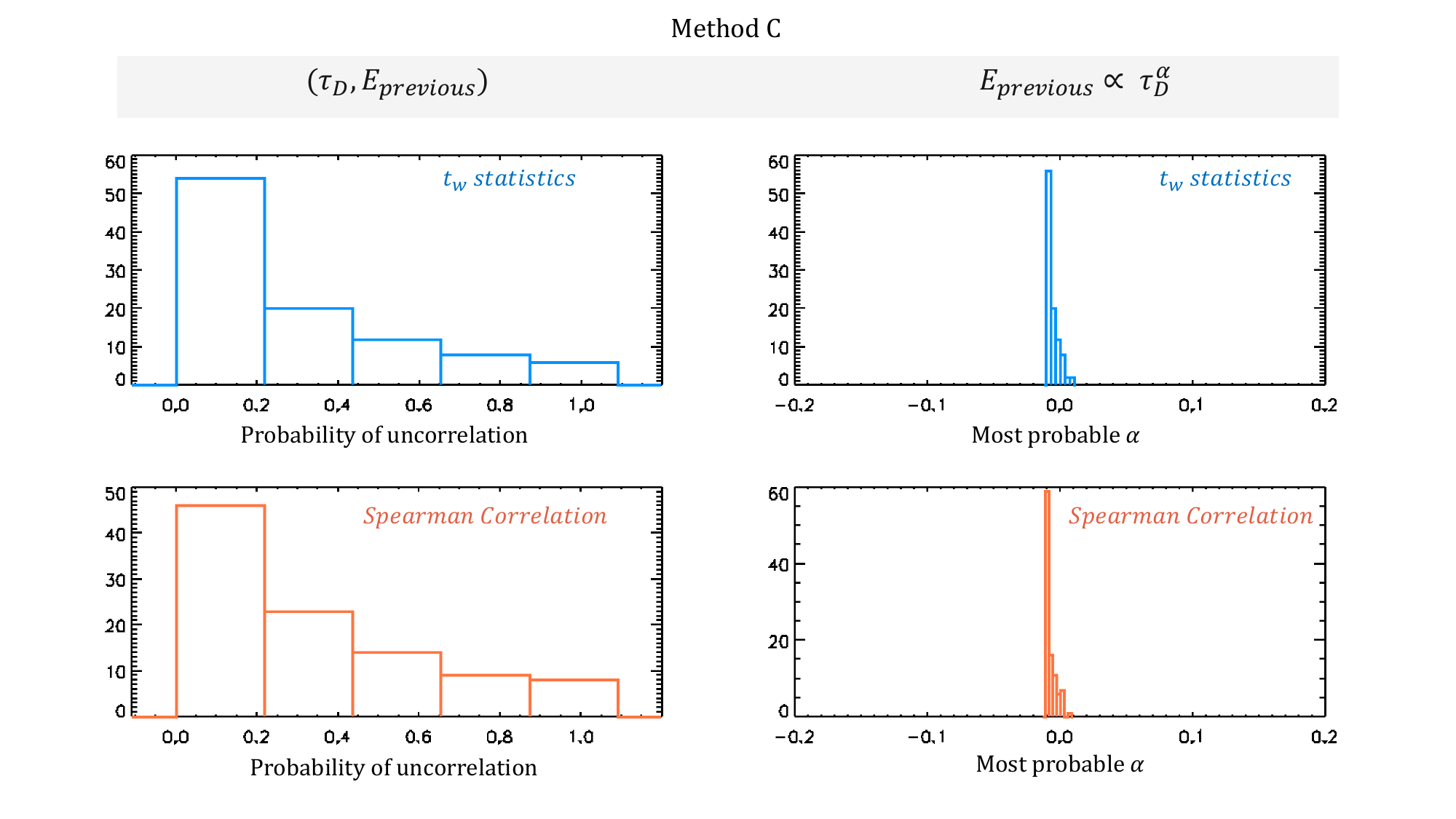}
 \includegraphics[width=0.95\textwidth,angle=0,page=6]{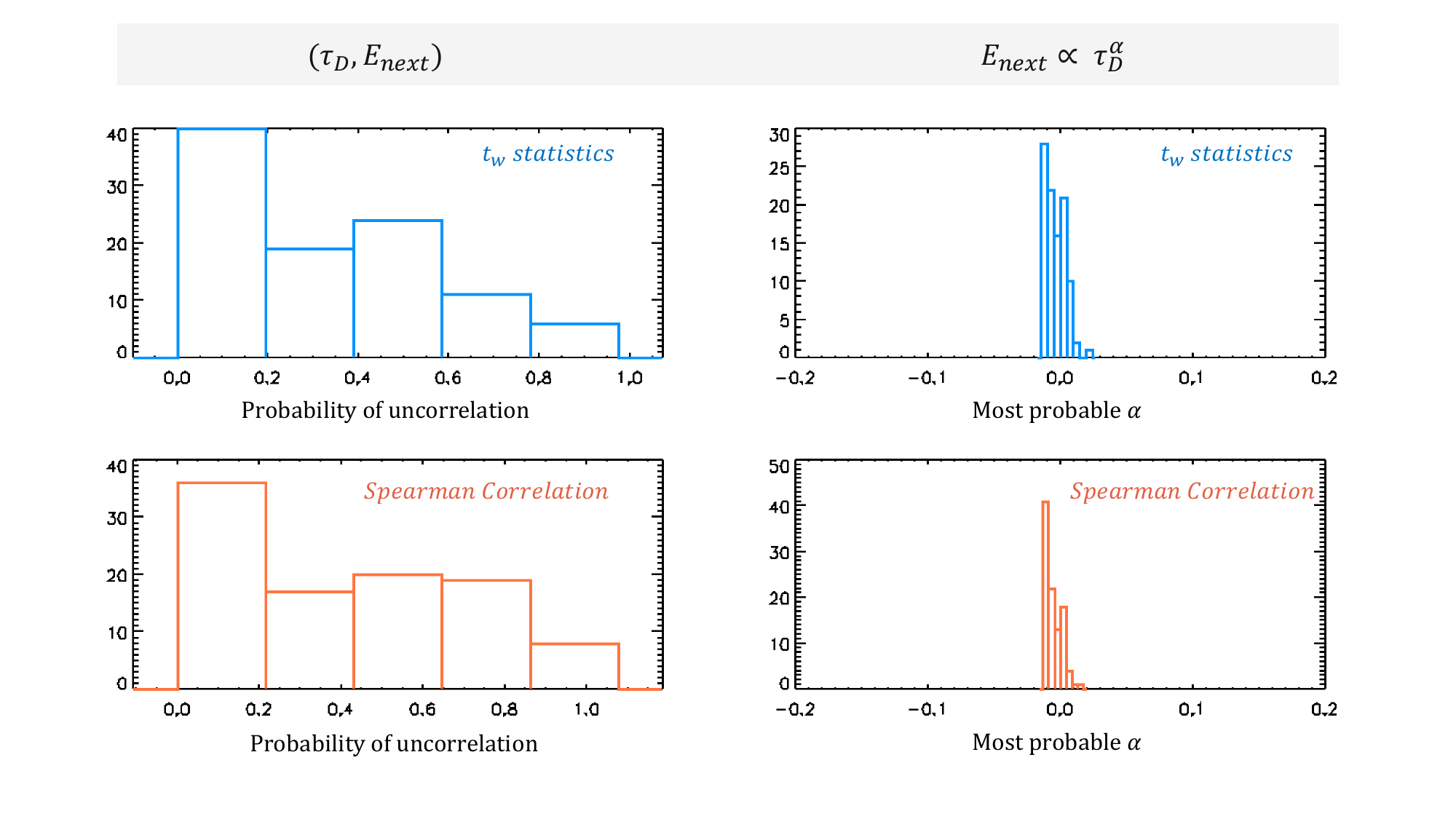}
 \caption{Statistical correlation results obtained via bootstrapping of the high-energy nanoflare samples identified using Method A, B and C. Left panels show the histograms of the probability of uncorrelation, while right panels display the most probable values of $\alpha$. While low p-values may imply potential correlations, the clustering of $\alpha$ values near zero indicates that no strong or meaningful dependence exists between $\tau_D$ and either $E_{\text{previous}}$ or $E_{\text{next}}$ for the high-energy nanoflares.}
  \label{high_energy}
  \end{figure*}

\bibliography{energy_delay.bib}
\end{document}